\documentclass[a4paper,11pt]{article}
\usepackage{pos}
\usepackage{amsmath}
\usepackage{color}
\usepackage{cancel}
\usepackage{ulem}

\title{Building Hadron Potentials from Lattice QCD with Deep Neural Networks}

\author*[a]{Lingxiao Wang}

\author[a]{Takumi Doi}

\author[a]{Tetsuo Hatsuda}

\author[a]{Yan Lyu}
\affiliation[a]{Interdisciplinary Theoretical and Mathematical Sciences Program (iTHEMS), RIKEN\\, Wako 351-0198, Japan}

\emailAdd{lingxiao.wang@riken.jp, doi@ribf.riken.jp, thatsuda@riken.jp, yan.lyu@riken.jp}

\abstract{In this study, we develop a deep learning method to learn hadronic interactions unsupervisedly from the correlation functions calculated in lattice QCD simulations. We present our approach of using deep neural networks to model the inter-hadron potentials that are learned from Nambu-Bethe-Salpeter (NBS) wave functions. This enables the incorporation of most general forms of potentials into the Schr\"odinger-type equation for detailed analysis of hadronic interactions.  Our results include validations with separable potentials, as well as the local and non-local potentials for the 
$\Omega_{ccc}-\Omega_{ccc}$ system. The neural networks accurately capture the essential features of these interactions, providing a reliable tool for predicting and analyzing hadron scattering properties, potentially bridging the experimental observables and lattice QCD data.
}

\FullConference{%
  The 41st International Symposium on Lattice Field Theory (Lattice 2024),\\
  July 28th - August 3rd, 2024\\
  the University of Liverpool, United Kingdom
}

\begin{document}
\maketitle

\section{Introduction}

Hadronic interactions play a crucial role in the formation of structures ranging from atomic nuclei to neutron stars~\cite{Epelbaum:2008ga,Baym:2017whm}. Traditional field-theoretical models based on the meson exchange picture have successfully described interactions between baryons at long distances. However, the short-range part of the interaction is less understood, likely due to the significant role of the quark-gluon structure at these scales. Lattice Quantum Chromodynamics (LQCD) provides a first-principles framework to study the hadronic interactions, allowing for a more fundamental and unifying approach by incorporating the dynamics of quarks and gluons. 

The HAL QCD method~\cite{Ishii:2006ec,Aoki:2008hh,Aoki:2009ji,Ishii:2012ssm} has been proposed to build effective potentials between hadrons from their spatial correlations, the equal-time Nambu-Bethe-Salpeter (NBS) amplitude, measured on the lattice, bridging the gap between LQCD and experimental data (see e.g.~\cite{ALICE:2020mfd}). The HAL QCD method has been applied to various baryon-baryon, meson-baryon and meson-meson systems; recently, the detailed anlysis of the doubly charmed tetraquark, $T_{cc}$, with $m_{\pi}=146$ MeV has also been reported~\cite{Lyu:2023xro}. Comprehensive reviews are available in Refs.~\cite{Aoki:2020bew,Aoki:2023qih}. In this method, the integral kernel of the integro-differential equation for the NBS wave function is treated as a non-local
potential between hadrons. In practice, the non-locality has been treated by the velocity expansion \`{a} la Okubo and Marshak~\cite{Okubo:1958qej} and the leading-order and the next-to-leading order potentials have been derived~\cite{HALQCD:2018gyl}. However, the truncation of the higher order terms of the velocity expansion as well as the fitting process of the potential with a few parameters bring additional uncertainties beyond the statistical and systematic errors of the original lattice data.

From the perspective of inverse problems, physics-driven deep learning offers more flexible solutions for constructing general potential functions from LQCD data~\cite{Shi:2021qri,Wang:2021jou,Zhou:2023pti}. In this study, we present a novel deep learning approach that directly models non-local hadronic interactions from LQCD correlation functions. Using deep neural networks to represent general potential functions, we train the model unsupervisedly with NBS wave functions. This approach provides greater flexibility compared to previous methods, which were often limited by specific assumptions. Moreover, explicit exchange symmetry is incorporated into the neural network to reduce the parameter space and regularize optimization. Validated on separable potentials, this method also successfully models the $\Omega_{ccc}-\Omega_{ccc}$ interactions within a general potential function.

\section{HAL QCD Method}

The equal-time Nambu-Bethe-Salpeter (NBS) amplitude $\phi_{\bf k}({\bf r})$, whose asymptotic behavior at large distances reproduces the scattering phase shift, plays an important role in the HAL QCD method.
A non-local potential $U({\bf r},{\bf r}^{\prime})$ 
for two baryons with an equal mass $m_B$ can be defined as ~\cite{Ishii:2006ec,Aoki:2008hh,Aoki:2009ji},  
\begin{equation}
    ({ E}_{\bf k}-{ H}_{0}){\phi}_{\bf k}({\bf r})=\int d^{3}\,r^{\prime}{ U}({\bf r},{\bf r}^{\prime}){\phi}_{\bf k}({\bf r}^{\prime}),\quad{ E}_{\bf k}=\frac{{\bf k}^2}{2m},\quad{H}_{0}=-{\frac{{\nabla}^{2}}{2m}},\quad m={\frac{m_{B}}{2}}.\label{eq:schor}
\end{equation}
Since all the elastic scattering states are governed by the same potential $U({\bf r},{\bf r}^{\prime})$, the time-dependent HAL QCD method~\cite{Ishii:2012ssm} takes full advantage of all the NBS amplitudes below the inelastic threshold  $\Delta E^* \sim \Lambda_{\rm QCD}$ by defining  so-called the $R$ correlator as $R(t, {\bf r})= \sum_{n}^{\infty}A_{n}\psi_{n}({\bf r})e^{-(\Delta W_{n})t}+O(e^{-(\Delta E^{*})t}),$ where $A_n$ is the overlapping factor, and $\Delta W_n = 2\sqrt{m^2_B + {\bf k}_n^2} - 2m_B$ with the relative momentum ${\bf k}_n$. The contributions from the inelastic states are exponentially suppressed when $t\gg (\Delta E^*)^{-1}$. In such condition, the $R$ correlator can be shown to satisfy following integro-differential equation~\cite{Ishii:2012ssm} as follows,
\begin{equation}
    \left\{\frac{1}{4m_{B}}\frac{\partial^{2}}{\partial t^{2}}-\frac{\partial}{\partial t}-H_{0}\right\}R(t,{\bf r})=\int d^{3}{\bf r^{\prime}}\,U({\bf r},{\bf r}^{\prime})R(t,{\bf r}^{\prime}).\label{eq:timeeq}
\end{equation}
The effective central potential in the leading order approximation of the velocity expansion, $U({\bf r},{\bf r}^{\prime}) = V(r)\delta({\bf r}-{\bf r^{\prime}})+\Sigma_{n=1}V_{2n}({\bf r})\nabla^{2n}({\bf r}-{\bf r^{\prime}})$, can be computed directly from,
\begin{equation}
   V(r) = \frac{1}{R(t,{\bf r})}\left\{\frac{1}{4m_{B}}\frac{\partial^{2}}{\partial t^{2}}-\frac{\partial}{\partial t}-H_{0}\right\}R(t,{\bf r}).
\end{equation}
Also, the higher-order terms of the velocity expansion $V_{2n}$ can be obtained by combining the information of the $R$ correlators obtained from different source operators or equivalently different weight factors $A_n$~\cite{HALQCD:2018gyl}.

\section{Physics-Driven Learning Hadron Potential}
\begin{figure}[!hbpt]
    \centering
    \includegraphics[width = 0.7 \textwidth]{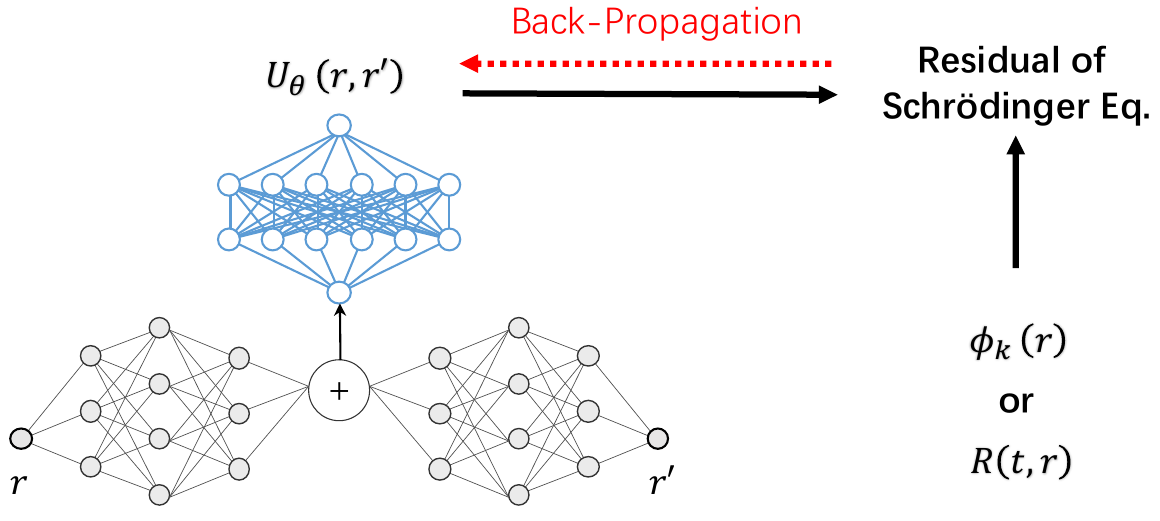}
    \caption{Symmetric deep neural network (SDNN) for representing potential functions. The gray colored neural networks are the same for inputs $r$ and $r^\prime$. The outputs of two different inputs are added in the latent layer, which is used as the input of the next layer neural network colored as blue. The final output represents $U_\theta(r, r^\prime)$.}
    \label{fig:symnn}
\end{figure}

To simplify without losing generality, we begin the reconstruction task using the wave function in the S-wave 
for systems of two identical particles. We design a parameter-sharing neural network to preserve the exchange symmetry in the non-local potential of such systems. Figure~\ref{fig:symnn} illustrates a sketch of the symmetric deep neural network (SDNN) used to represent the potential $U_\theta(r, r^\prime)$. The inputs are $(r, r^\prime)$, and the output from the parameter-sharing network is $f(r)$, which is then combined with $f(r^\prime)$ as input to the next layer. The final output of the network is given by $U_\theta(r, r^\prime) \equiv g(f(r) + f(r^\prime))$, where $g(x)$ and $f(x)$ are two distinct neural networks, and $\{ \theta \}$ indicates all trainable parameters inside the neural network~\footnote{In a highly simplified case, the identical mapping $f(x) \equiv x$ could be used, but this would limit the representation ability of neural networks.}.

Given the wave function $\phi_{k}({r})$ and the energy $E_k$, the potential $U_\theta({r},{r}^{\prime})$ can be determined by minimizing the following loss function,
\begin{equation}
    \mathcal{L}_k=\sum_k \sum_r \,\left[\left({ E}_{k}-{ H}_{0}\right){\phi}_{k}({r})-\int 4\pi {r^{\prime}}^2 d\,r^{\prime}U_\theta({r},{r}^{\prime}){\phi}_{k}({r}^{\prime})\right]^{2}. \label{eq:loss_k}
\end{equation}
Furthermore, given the correlation function $R(t,r)$, the potential $U_\theta(r ,r')$ can be determined by minimizing the following loss function,
\begin{equation}
    \mathcal{L}_t=\sum_t \sum_r \left\{\frac{1}{4m_{B}}R_{tt}(t,r) - R_{t}(t,r) + \frac{1}{m_B }R_r(t,r)- \int 4\pi {r^{\prime}}^2 dr^{\prime}\,U_\theta(r,r^{\prime})R(t,r^{\prime}) \right  \}^2, \label{eq:loss_t}
\end{equation}
where $R_{tt}(t,r) \equiv \partial_t^2 R(t,r)$, $R_{t}(t,r) \equiv \partial_t R(t,r)$, and $R_{r}(t,r) \equiv \nabla^2 R(t,r)$. The loss function is nothing but the residual of Eq.~\eqref{eq:timeeq}. 

To introduce the physical constraint as a regularization, we adopt the asymptotic behaviour of the hadron-hadron interaction, $\lim_{r,r\rightarrow \infty} U(r,r^{\prime}) = 0$, as the regularization loss function,
\begin{equation}
    \mathcal{L}_r =  \sum_{n,m}^N U_\theta(r_n,r^\prime_m) ^2, \quad r_n>\tilde{R}, \quad  r^{\prime}_m >\tilde{R},
\end{equation}
where $\tilde{R}$ is a cutoff for indicating there is zero potential. The total loss function becomes, $\mathcal{L}\equiv \mathcal{L}_{\text{data}} + \mathcal{L}_r$. As Figure~\ref{fig:symnn} shows, the wave function $\phi_{k}({r})$ (correlation function $R(t,r)$) and potential function $U_\theta(r ,r')$ are used to compute the residual, and further used to calculate the gradients to parameters of neural networks. The gradient-based algorithm, back-propagation (BP) method~\cite{bishop2023deep}, is applied to optimize the neural network parameters $\{ \theta \}$ by,
\begin{equation}
    \theta_{i+1} \rightarrow \theta_i + \frac{\partial \mathcal{L}}{\partial U_\theta(r,r^\prime)} \frac{\partial U_\theta(r,r^\prime)}{\partial \theta},
\end{equation}
where the index $i$ labels the time-step in optimization process.

\section{Numerical Results}

\subsection{Separable Potential}
We start from a solvable potential, the separable potential~\cite{Aoki:2020bew} used as a toy model for demonstration. The definition of the radial potential is,
\begin{equation}
    U({\bf r},{\bf r}^{\prime})\,\equiv\,\omega\nu({\bf r})\nu({\bf r}^{\prime}),\;\;\;\;\;\nu({\bf r})\equiv e^{-\mu r},
\end{equation}
where $\omega, \mu$ are parameters. The S-wave solution of the Schr\"odinger equation with this potential is given exactly by,
\begin{equation}
    {{\phi_{k}^{0}(r)}} {{=\displaystyle\frac{e^{i\delta_{0}(k)}}{k r}\left[\sin(k r+\delta_{0}(k))-\sin\delta_{0}(k)e^{-\mu r}\left(1+\displaystyle\frac{r(\mu^{2}+k^{2})}{2\mu}\right)\right],}}
\end{equation}
where, $k\cot\delta_{0}(k)=-\frac{1}{4\mu^{2}}\left[2\mu(\mu^{2}-k^{2})-\frac{3\mu^{2}+k^{2}}{4\mu^{3}}(\mu^{2}+k^{2})^{2} +\frac{(\mu^{2}+k^{2})^{4}}{8\pi m\omega} \right]$. As a numerical example, we take $\mu=1.0$, $\omega=-0.017 \mu^4$ and $m=3.30 \mu$ and $R = 2.5/\mu$, the physics unit is chosen as $\mu$. 
When setting ${\varphi}^0_{ k}({ r}) / r\equiv {\phi}^0_{ k}({ r})$, in the spherically symmetric case, the radial euqation will be derived as,
\begin{equation}
    \left( \frac{d^2}{dr^2} + k^2 \right) {\varphi}^0_{ k}(r) = 8\pi mr \int r^{\prime} dr^{\prime} U(r, r^{\prime}) {\varphi}^0_{ k}(r^{\prime}).
\end{equation}

\begin{figure}[!hbpt]
    \centering
    \includegraphics[width=1.\linewidth]{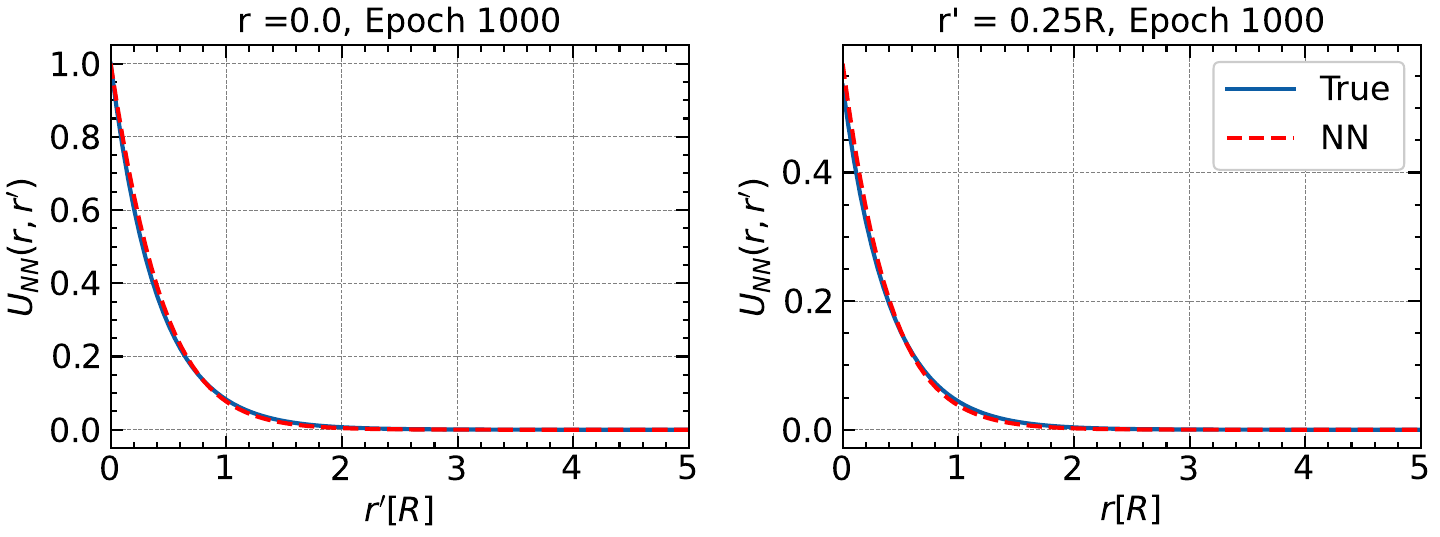}
    \caption{Reconstructed separable potentials from neutral network (NN) and the ground truths.}
    \label{fig:pot-sep}
\end{figure}

A practical setup for preparing wave functions involves using momentum values $k = [0.01, 1.0]$ with $N_k = 10$ and radial distances $r = [0.01, 5R]$ with $N_r = 100$. A total of 1000 data points are employed to minimize the loss function in Eq.~\eqref{eq:loss_k} for an optimal potential $U_{NN}(r, r^\prime) \equiv \omega U_\theta(r, r^\prime)$. The SDNN configuration, as illustrated in Fig.~\ref{fig:symnn}, consists of two identical network paths that process inputs $r$ and $r^\prime$ through a series of linear transformations (structured as $1 \rightarrow 64 \rightarrow 16$) with LeakyReLU activations. These outputs are additively combined to enforce symmetry, and the merged feature is passed through a final linear layer (structured as $16 \rightarrow 1$). A Softplus activation function is applied to the output to ensure smooth, positive predictions. The reconstructed potential is shown in Fig.~\ref{fig:pot-sep}. Regularization is achieved by imposing the asymptotic behavior $U_\theta(r > \tilde{R}, r^\prime > \tilde{R}) = 0$, where $\tilde{R} = 2R$. After 1000 epochs of training, the symmetric deep neural network successfully recovers the ground truth potential functions.

\subsection{$\Omega_{ccc}-\Omega_{ccc}$ Interaction}

\begin{figure}[!hbpt]
    \centering
    \includegraphics[width=0.5\linewidth]{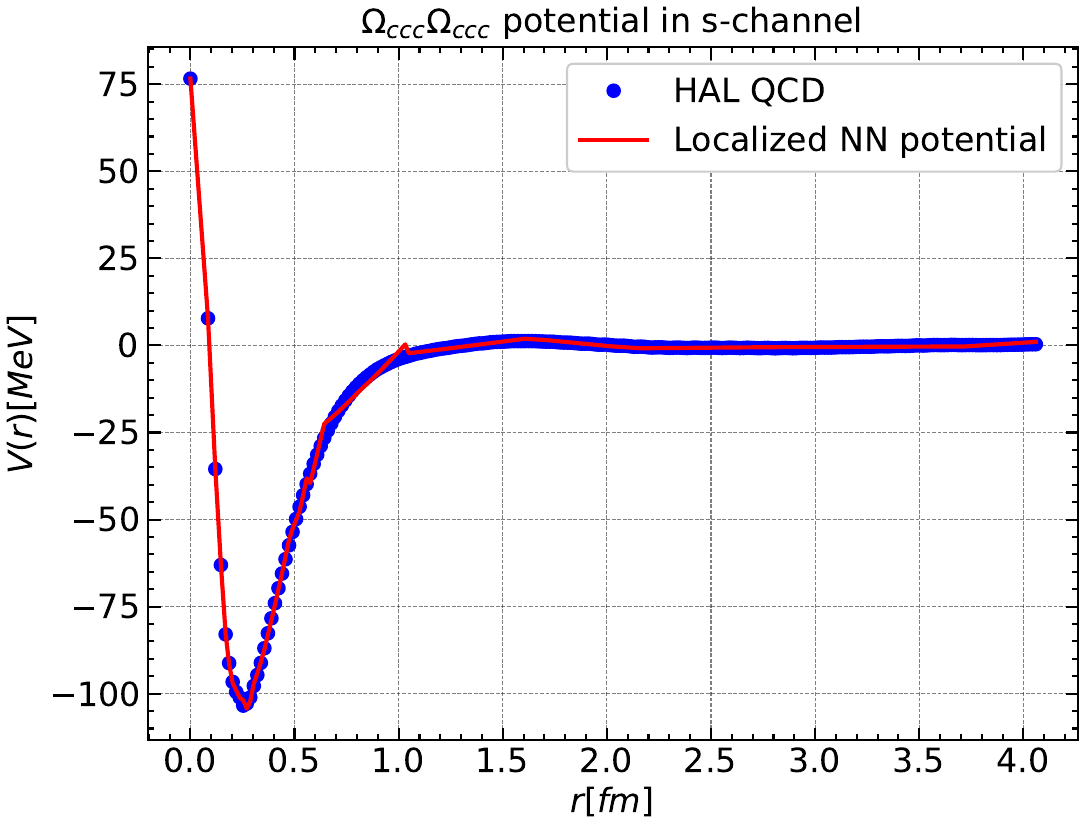}
    \caption{Localized neural network potential $V_\theta (r) $ and the previous HAL QCD reconstruction for $\Omega_{ccc}-\Omega_{ccc}$ interactions~\cite{Lyu:2021qsh}.}
    \label{fig:pot-oo-1}
\end{figure}

For the second demonstration, we focus on the $\Omega_{ccc}-\Omega_{ccc}$ system, which has been extensively studied in current research~\cite{Lyu:2021cte,Lyu:2021qsh}. The gauge configurations used in this work employ a $(2+1)$-flavor setup on a $96^4$ lattice with the Iwasaki gauge action at $\beta = 1.82$. The lattice spacing is approximately $a \approx 0.0846 \, \text{fm}$ ($a^{-1} \approx 2.333 \, \text{GeV}$), with pion and kaon masses set to $m_\pi \approx 146 \, \text{MeV}$ and $m_K \approx 525 \, \text{MeV}$, respectively. The interpolated mass for the $\Omega_{ccc}$ is $m_{\Omega_{ccc}} \approx 4796 \, \text{MeV}$. Further details on the lattice setup are available in Ref.~\cite{Lyu:2021qsh}. 

\begin{figure}[!hbpt]
    \centering
    \includegraphics[width=0.5\linewidth]{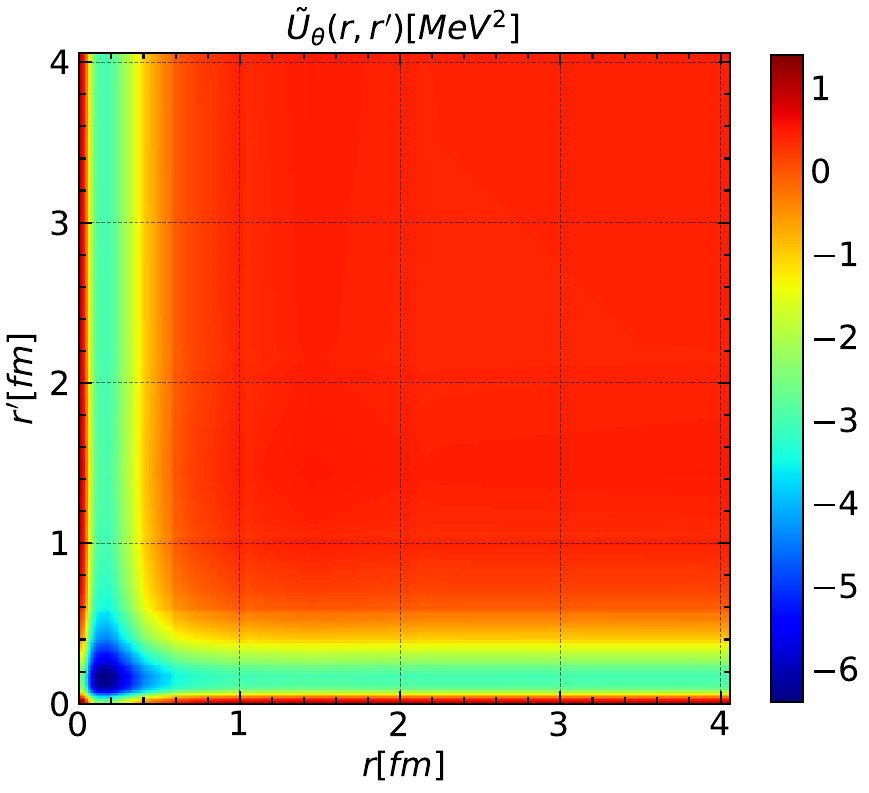}
    \caption{Non-local neural network potential for $\Omega_{ccc}-\Omega_{ccc}$.
    }
    \label{fig:pot-oo-2}
\end{figure}

In this setup, the symmetric deep neural network (SDNN) path is deeper than in the previous case, following the structure ($1\rightarrow 32\rightarrow64\rightarrow128\rightarrow64\rightarrow32\rightarrow 16$), with an ELU activation applied to the final output. Other configurations remain the same as the previous case. We trained the model using $R(t = 26, r)$ correlation data of the $\Omega_{ccc}-\Omega_{ccc}$ system calculated in lattice QCD simulations, where $t = 25, 27$ are only used for computing $R_{tt}, R_t$. A regularization is applied by enforcing the asymptotic behavior $U_\theta(r > 3\,\text{fm}, r' > 3\,\text{fm}) = 0$. After 5000 epochs, the localized potential, $V_\theta (r)$, defined as $V_\theta (r) \equiv ({\sum_{r'} \Delta r'\,\tilde{U}_\theta(r,r')R(t,r')})/{R(t,r)}$ with $\tilde{U}_\theta(r,r')\equiv 4\pi {r^{\prime}}^2{U}_\theta(r,r'), \Delta r' = 0.2 a$, is compared to the HAL QCD reconstruction. The comparison is shown in Fig.~\ref{fig:pot-oo-1}, and the non-local 3D potential function is demonstrated for the first time in Fig.~\ref{fig:pot-oo-2}.

\section{Summary}
In this study, we develop a deep learning approach to learn hadronic interactions in an unsupervised manner from the  correlation functions calculated in lattice QCD. Specifically, we use deep neural networks to model potential functions derived from Nambu-Bethe-Salpeter (NBS) wave functions. Our framework is validated by successfully reconstructing a solvable separable potential. Further numerical results demonstrate that the $\Omega_{ccc}-\Omega_{ccc}$ potential can be directly constructed from $R$ correlators, achieving consistency with previous HAL QCD results. Notably, the full non-local potential $U(r,r')$  is also reconstructed for the first time. Future work includes calculating phase shifts and applying joint learning with heavy-ion collision data to better understand hadron-hadron interactions from two complementary sources.

\section*{Acknowledgement}
We thank the members of HAL QCD Collaboration, Drs. Shuzhe Shi, Jiaxing Zhao, Kai Zhou for helpful discussions. The lattice QCD measurements have been performed on HOKUSAI supercomputers at RIKEN. T.D., T.H. and Y.L. were supported by Japan Science and Technology Agency (JST) as part of Adopting Sustainable Partnerships for Innovative Research Ecosystem (ASPIRE), Grant Number JPMJAP2318. This work was also supported by JSPS (JP19K03879,JP23H05439) and MEXT (JPMXP1020230411). L.W. also thanks the National Natural Science Foundation of China(No.12147101) for supporting his visit in Fudan University when preparing this work.

\bibliographystyle{apsrev4-1}
\bibliography{nnhal}

\end{document}